\documentclass{emulateapj}
\usepackage{apjfonts}
\usepackage{natbib}
\usepackage{amsmath}
\usepackage{graphicx}
\def \ni {\noindent}
\def\lsim{\mathrel{\rlap{\lower4pt\hbox{\hskip1pt$\sim$}}
    \raise1pt\hbox{$<$}}}                
\def\gsim{\mathrel{\rlap{\lower4pt\hbox{\hskip1pt$\sim$}}
    \raise1pt\hbox{$>$}}}                

\def\fmls{$M/L_V$ }
\def\mbh{$M_{\bullet}$}
\def\mbhs{$M_{\bullet}$ }

\def\apjl{ApJL}
\def\aj{AJ}
\def\mnras{MNRAS}
\def\apj{ApJ}
\def\araa{ARA\&A}
\def\nat{Nature}

\def\apjs{ApJS}

\shorttitle{The Dark Matter Density Profile of the Fornax Dwarf}
\slugcomment{{\sc Accepted to ApJ:} 29 November 2011} 

\defcitealias{nav96}{NFW}

\begin{document}
\title{The Dark Matter Density Profile of the Fornax Dwarf}
\author{John R. Jardel and Karl Gebhardt}

\affil{Department of Astronomy, University of Texas at
Austin, 1 University Station C1400, Austin, TX 78712;\\
jardel:gebhardt@astro.as.utexas.edu}

\begin{abstract}

We construct axisymmetric Schwarzschild models to measure the mass profile of 
the local group dwarf galaxy Fornax.  These models require no assumptions to 
be made about the 
orbital anisotropy of the stars, as is the case for commonly used Jeans 
models.  We test a variety of parameterizations
of dark matter density profiles and find cored 
models with uniform density 
$\rho_c=(1.6 \pm 0.1) \times 10^{-2} \, M_{\odot} \text{ pc}^{-3}$
fit significantly better than the cuspy halos predicted by cold dark matter
simulations.
We also construct models with an intermediate-mass black hole, but are unable
to make a detection.  We place a $1$-$\sigma$ upper limit on the mass of a
potential intermediate-mass black hole at \mbh$\leq 3.2 \times 10^4 \, M_{\odot}$.

\end{abstract}


\section{Introduction}

Low-mass galaxies provide a unique testing ground for predictions of the 
cold dark matter (CDM) paradigm for structure formation, since they generally 
have a lower fraction of baryons than massive galaxies.  These galaxies 
allow for a more direct 
measurement of the underlying dark matter potential, as the complicated effects
of baryons on the dark matter are less pronounced.  A particularly testable
prediction of CDM is that all galaxies share a universal dark matter density 
profile, characterized by a cuspy inner power law $\rho \propto r^{-\alpha}$
where $\alpha=1$ (\citealt{nav96}, hereafter \citetalias{nav96}).  Many authors
have investigated low-mass spirals and found, in contrast to the predictions
of CDM, dark matter density profiles with a flat inner core of slope
$\alpha=0$ \citep{bur95,per96,deb01,bla01,sim05}.  This has launched the 
debate known as the core/cusp controversy.  

A number of other studies have investigated the mass content of dwarf 
spheroidal galaxies (dSphs).  
\citet{gil07} give a comprehensive review of recent attempts
to constrain the inner slope of their dark matter profiles with Jeans modeling 
(\citealt{jea19}; \citealt{bt87}, chapter 4).
When significant, cored 
profiles are preferred for all dSphs modeled (\citealt{gil07}, and references
therein).

These results, however, are subject to a major caveat of Jeans modeling; it is 
complicated by the effect of stellar velocity anisotropy.  
Models fit to the line-of-sight component of the velocity dispersion, but
anisotropy can severely
affect the modeling of enclosed mass.  Therefore, additional assumptions must
be made.  The studies presented in \citet{gil07}
assume spherical symmetry and isotropy.  \citet{eva09} show that 
a weakness of Jeans modeling is that given these assumptions combined with 
the cored light profiles observed in dSphs, the Jeans equations do not allow
solutions with anything other than a cored dark matter profile.

\citet{wal09b} construct more sophisticated models and attempt to parameterize
and fit for the anisotropy.  As a result, preference for cored profiles 
becomes 
model-dependent.  They therefore are unable to put significant constraints
on the slope of the dark matter profile.  This highlights the main problem
with Jeans modeling---it is highly dependent on the assumptions made.

Distribution function models are more general than Jeans models,
and progress has been made applying them 
to a number of dSph systems \citep{kle02,wu07,amo11}.
Nevertheless these models still make strong assumptions such as
spherical symmetry or isotropy, and models that do fit for anisotropy do so
without using the information about the stellar orbits contained in the 
line-of-sight velocity distributions (LOSVDs).

We employ a fundamentally different modeling technique, known as 
Schwarzschild modeling, that allows us to use this information to  
self-consistently calculate both the enclosed mass and orbital anisotropy.
Schwarzschild modeling is a mature industry, but one that has seldom
been applied to the study of dSph galaxies (see \citealt{val05}).

In addition to being well-suited for measuring dark matter profiles, 
Schwarzschild modeling has often been used to search
for black holes at the centers of galaxies.
Another unresolved issue relevant to the study of dSphs is whether they host
an intermediate-mass black hole (IMBH).  In a hierarchical merging scenario, 
smaller galaxies are thought to be the
building blocks of larger galaxies.  It is thought that all massive galaxies
host a supermassive black hole (SMBH) at their center, therefore it is logical
to believe that their building blocks host smaller IMBHs.  Evidence for these 
IMBHs is scarce, however, and dynamical detections are even scarcer.  
The closest and lowest mass example of a dynamical 
measurement is an upper limit on the local group dSph NGC~205 of
\mbh$\leq 2.2 \times 10^4 \, M_{\odot}$ \citep{val05}.  Black holes in this
mass range can provide constraints on theories of black hole growth and
formation.  The two most prominent competing theories of nuclear black hole
formation are direct collapse of primordial gas \citep{ume93,eis95,beg06}
or accretion onto and mergers of seed black holes resulting from the collapse
of the first stars \citep{vol05}.

In this paper we present 
axisymmetric, three-integral Schwarzschild models in an effort to determine
the inner slope of the dark matter density profile as well as the orbit 
structure
of the Fornax dSph.  We also investigate the possibility of a central
IMBH.  We assume a distance of 135 kpc to Fornax \citep{ber00}.


\section{Data}

To construct dynamical models, we require a stellar light profile as well as
stellar kinematics in the form of LOSVDs.  We use published data for both
the photometry and kinematics, and describe the steps taken to convert this
data into useful input for our models.

\begin{figure}[t]
\includegraphics[width=9cm]{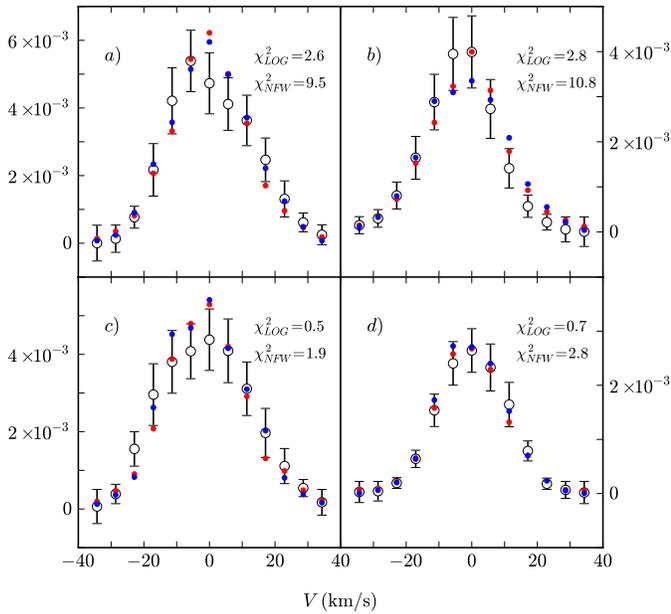}
\caption{Line-of-sight velocity distributions of four bins.
Open circles with error bars are the data.  Over-plotted are the model values
for the best-fitting cored model (red) and NFW model (blue).  Bins are located
at: (a) $R=297\arcsec$, $\theta=18^{\circ}$ (b) $R=550\arcsec$,
$\theta=18^{\circ}$ (c) $R=1008\arcsec$, $\theta=45^{\circ}$
(d) $R=2484\arcsec$, $\theta=45^{\circ}$.  Quoted $\chi^2$ values are 
un-reduced.
\label{losvd}}
\end{figure}

\subsection{Stellar Density}

To determine the stellar density, we use a number density profile from
\citet{col05} extending to $4590\arcsec$.  We linearly 
extrapolate the profile out to $6000\arcsec$---a physical radius of 3.9 kpc at
our assumed distance.  We also extrapolate the profile inwards at constant
density from $90\arcsec$ to $1\arcsec$. 

To convert to a more familiar surface brightness
profile we apply an arbitrary zero-point shift in log space, adjusting this 
number so that
the integrated profile returns a luminosity consistent with the value listed
in \citet{mat98}.  
Adopting an ellipticity of $e=0.3$ \citep{mat98}, we deproject under
the assumption that surfaces of constant luminosity are coaxial spheroids 
\citep{geb96}, and for an assumed inclination of $i=90^{\circ}$.

\subsection{Stellar Kinematics}

We derive LOSVDs from individual stellar velocities published in 
\citet{wal09}.  The data contain heliocentric radial velocities and 
uncertainties with a membership probability for 2,633
Fornax stars.  Most of these are single-epoch observations, however some are
multi-epoch.  Stars that have more than one observation are averaged, weighted
by their uncertainties.  After making a cut in membership probability at 
90\%, we are left with 2,244 stars.
Although a significant number of stars observed may be in binary or multiple
systems, simulations have shown that such systems are unlikely to affect 
measured dispersions \citep{har96,ols96,mat98}.

We adopt a position angle $PA=41^{\circ}$ \citep{wal06}.  We assume symmetry
with respect to both the major and minor axes and fold the data along each
axis.  To preserve any possible rotation, we switch the sign of the velocity
whenever a star is flipped about the minor axis.

\begin{figure}[t]
\includegraphics[width=9cm]{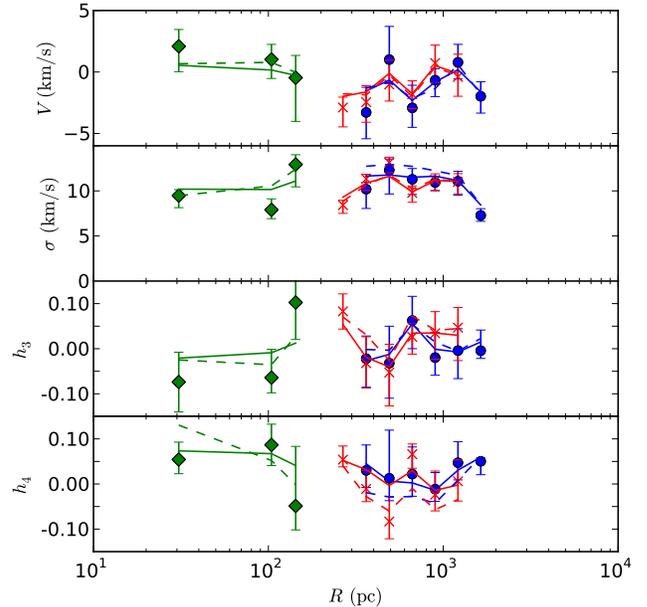}
\caption{Gauss-Hermite moments for stars near the major axis (blue),
minor axis (red), and averaged over all angles (green).  Solid lines 
correspond to the best-fit model with a cored dark matter halo,
dashed lines are for the best-fit model with a NFW halo.
\label{moments}}
\end{figure}

The transverse motion of Fornax contributes a non-negligible line-of-sight
velocity to stars, particularly those at large galactocentric radius.  
Using the equations in Appendix A of \citet{wal08}, we correct for this effect.
We adopt values for the proper motion of 
$(\mu_{\alpha},\mu_{\delta})=(47.6, -36.0)\text{ mas century}^{-1}$
\citep{pia07} and assume the heliocentric radial velocity of Fornax is
$53.3\text{ km s}^{-1}$ \citep{pia02}.

We divide our meridional grid into 20 radial bins, equally spaced in
approximately $\log \,r$ 
from 1\arcsec to 5000\arcsec.  There are 5 angular bins spaced equally
in sin $\theta$ over 90$^{\circ}$ from the major to the minor axis
\citep{geb00,sio09}.  From the positions of the folded stellar velocity data,
we determine the best binning scheme so that each grid cell contains at least 
25 stars from which to recover the LOSVD.  Our first bin with enough stars to
meet this criterion is centered  at 47\arcsec, and the last bin is centered 
at 2500\arcsec.  We therefore have two-dimensional kinematics coverage over 
the radial range 47\arcsec-2500\arcsec (30pc - 1.6 kpc).  At small radii the 
number density of stars with velocity measurements is low, thus our 
central LOSVDs have higher uncertainty compared to those at larger radii.

\begin{deluxetable*}{llllllllll}
\centering
\tablecaption{Best-Fit Model Parameters}
\tablewidth{0pt}
\tablehead{
	\colhead{DM Profile} & \colhead{$\chi^2$} & 
	\colhead{$\frac{M}{L_V}$} & 
	\colhead{c} & \colhead{$r_s$ (kpc)} & 
	\colhead{$\rho_c$ ($M_{\odot} pc^{-3}$)} & 
	\colhead{$M_{\bullet}$($M_{\odot}$)} & \colhead{$N_{model}$}}
\startdata
NFW & 239.8 & $1.3 \pm 0.6$ & $4.1 \pm 0.26$ & $11.7 \pm 1.4$ 
        & --- & --- & 3124 \\
Log & 162.6 & $1.5 \pm 0.5$ & --- & --- 
& $1.6 \pm 0.1 \times 10^{-2}$ &  --- & 4319 \\
Log & 162.6 & $1.6 \pm 0.2$ & --- & --- &
	 $1.6 \pm 0.1 \times 10^{-2}$ & $\leq 3.2 \times 10^4$ & 3423 \\
\enddata
\tablecomments{Best-fit parameters for NFW, and cored logarithmic
dark matter halos.  $\chi^2$ is un-reduced, the number of degrees of freedom 
are the same for each model.  Model parameters and $1$-$\sigma$ uncertainties 
are quoted.  $N_{model}$ lists the number of models run
for the corresponding parameterization.}

\label{restab}
\end{deluxetable*}
\vskip 20pt

Within each grid cell, we calculate the LOSVD from discrete stellar velocities
by using an adaptive kernel density estimate adapted from \citet{sil86} and
explained in \citet{geb96}.  We estimate the $1-\sigma$ uncertainties in the
LOSVDs through bootstrap resamplings of the data \citep{geb96,geb09}.
The bootstrap generates a new sample from the data itself by randomly picking $N$
data points, where $N$ is the total number of stars in a given bin, allowing
the same point to be chosen more than once.
We then estimate the LOSVD from that realization and repeat the procedure $300$
times.  The $68\%$ confidence band on the LOSVDs corresponds to the 
$68\%$ range of the realizations.
We compare the velocity dispersion as measured by the LOSVDs with the 
biweight scale (i.e., a robust estimate of the standard deviation, see 
\citealt{bee90}) of the individual velocities and note good 
agreement.

Figure \ref{losvd} plots the LOSVDs of four bins.  Rather than parameterizing
these LOSVDs with Gauss-Hermite moments, our models instead fit directly to
the LOSVDs to constrain the kinematics of the galaxy.  However, we do fit
Gauss-Hermite moments for plotting purposes only.  These data are presented in
Figure \ref{moments} for stars that have been grouped into bins near the
major axis (blue) and minor axis (red).  Near the center of the galaxy 
the density of stars with kinematics is sparse, so we therefore
group stars into annular bins covering all angles (green).  
We estimate the $1$-$\sigma$ uncertainties of the Gauss-Hermite moments
by fitting to each of the 300 realizations calculated during the 
bootstrap discussed above.  The error bars plotted contain 68\% of the 
300 realizations.

\section{Dynamical Models}

The modeling code we use is described in detail in
\citet{geb03},\citet{tho04,tho05}, and \citet{sio09} and is based on the 
technique of orbit superposition \citep{sch79}.  Similar axisymmetric codes 
are described in \citet{rix97,vdm98,cre99,val04} while \citet{vdb08} present
a fully triaxial Schwarzschild code.  Our code begins by choosing a
trial potential that is a combination of the stellar density, dark matter 
density, and 
possibly a central black hole.  We then launch $\sim 15,000$ orbits carefully
chosen to uniformly sample the isolating integrals of motion.  In an
axisymmetric potential, orbits are restricted by three isolating integrals
of motion, $E$, $L_z$, and the non-classical ``third integral'' $I_3$.
As it is not possible to calculate $I_3$ a priori, we use a carefully designed
scheme to systematically sample $I_3$ for each pair of $E$ and $L_z$
 \citep{tho04,sio09}.  Orbits are integrated for many dynamical times, and each 
orbit is given a weight $w_i$.   We find the combination of
$w_i$ that best reproduces the observed LOSVDs and light profile via a 
$\chi^2$ minimization subject to the constraint of maximum entropy 
\citep{sio09}.

We run models by varying 3 parameters---the stellar \fmls and two 
parameters specifying the dark matter density profile.  Some models are also 
run with a central black hole whose mass is varied in addition to the other 
3 model parameters.  Each model is assigned a value of $\chi^2$ and we 
identify the best-fitting model as that with the lowest $\chi^2$.  We 
determine the
$68\%$ confidence range on parameters by identifying the portion of their 
marginalized $\chi^2$ curves that lie within $\Delta \chi^2=1$ of the overall
minimum.

\subsection{Model Assumptions}

Our trial potential is determined by solving Poisson's equation for an assumed
trial density distribution.  On our two-dimensional polar grid, this takes the 
form:

\begin{equation}
\rho(r,\theta)= \frac{M}{L} \nu(r,\theta) + \rho_{DM}(r)
\label{denseq}
\end{equation}

\ni where $M/L$ is the stellar mass-to-light ratio, assumed constant
with radius, and $\nu(r,\theta)$ is the
unprojected luminosity density.  The assumed dark matter profile $\rho_{DM}(r)$
is discussed below.  For simplicity, we assume Fornax is edge-on in all our
models.

\subsection{Dark Matter Density Profiles}

We parameterize the dark matter halo density
with a number of spherical density profiles.  We use NFW halos:

\begin{equation}
\rho_{DM}(r)=\frac{200}{3} \frac{A(c)\rho_{crit}}{(r/r_s)(1+r/r_s)^2} 
\end{equation}

\ni where
\begin{equation*}
A(c)=\frac{c^3}{\ln(1+c)-c/(1+c)}
\end{equation*}

\ni and $\rho_{crit}$ is the present critical density for a closed universe. 
The two
parameters we fit for are the concentration $c$ and scale radius $r_s$.  
We also use halos derived from the logarithmic potential: 

\begin{equation}
\rho_{DM}(r)=\frac{V_c^2}{4 \pi G} \frac{3r_c^2+r^2}{(r_c^2+r^2)^2}
\end{equation}

\ni These models feature a flat central core of density 
$\rho_c = 3 V_c^2 / 4 \pi G r_c^2$ for $r \lsim r_c$ and an 
$r^{-2}$ profile for $r>r_c$.   We fit for
$V_c$ and $r_c$, the asymptotic circular speed at $r=\infty$ and core radius
respectively.  We run over 10,000 models with only three distinct
parameterizations: NFW halos, and logarithmic models with and without 
an IMBH.

\section{Results}

\begin{figure*}[t]
\centering
\includegraphics[width=15cm]{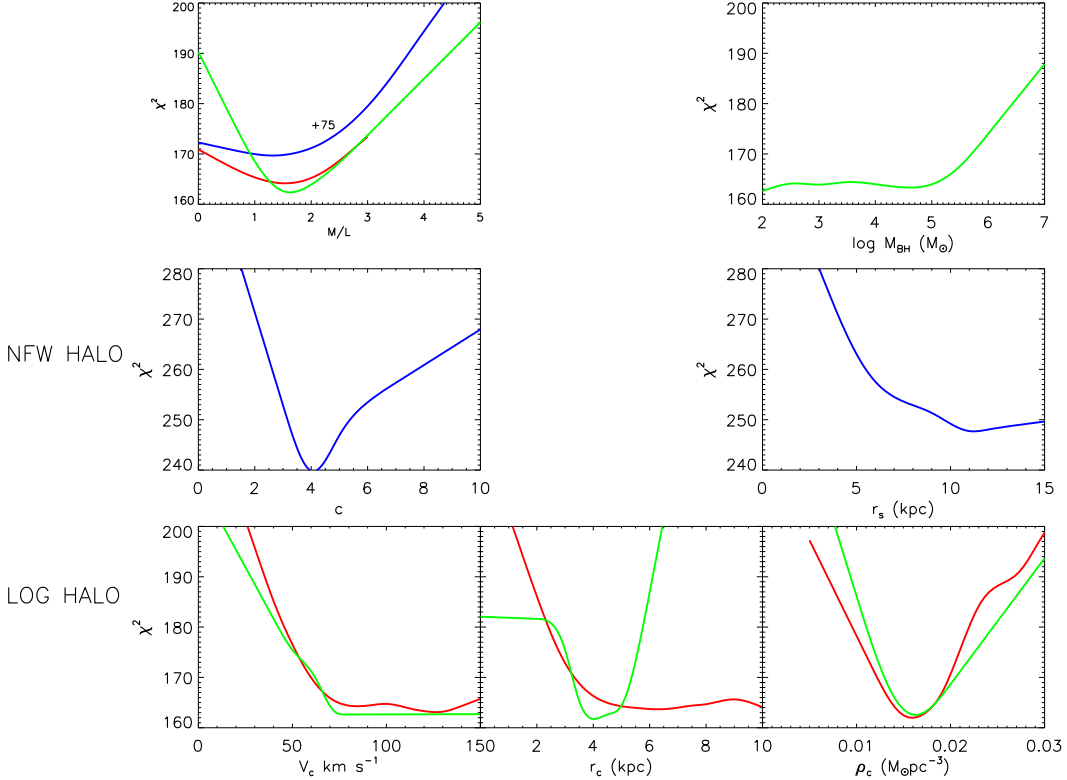}
\caption{$\chi^2$ curves for all parameterizations of the mass
profile.  NFW halos (blue) are parameterized by concentration $c$ and scale 
radius $r_s$.  Logarithmic halos with an IMBH (green) and without (red)
are specified by $V_c$ and $r_c$.  We also plot core density 
$\rho_c=3V_c^2/4\pi G r_c^2$ as it is the controlling parameter over the radial
range of our models.   We fit for stellar 
\fmls in all models (upper left panel).  NFW models have much higher 
$\chi^2$ and are scaled down by 75 to fit on the same axis.  Black hole mass
for logarithmic halos with an IMBH (green) is plotted in the upper right panel.
Note the apparent minimum in $r_c$ for logarithmic halos with an IMBH is due 
to incomplete parameter sampling.
\label{chi2res}}
\end{figure*}

We find significant evidence for cored logarithmic dark matter density profiles.
These models are preferred at the $\Delta \chi^2=77$ level when compared to
models with an NFW halo, a highly significant result.  
Perhaps more convincingly, the values for the concentration preferred by our
models are around $c=4$.  Only relatively recently formed structures
like galaxy clusters are expected to have concentrations this low
\citepalias{nav96}.

Table \ref{restab} summarizes the results of our models, 
while Figures \ref{losvd} and \ref{moments} illustrate the preference for
cored models
over models with an NFW halo in fitting to the kinematics.  We stress again 
that
LOSVDs like those plotted in Figure \ref{losvd} are the kinematic constraint,
and not the Gauss-Hermite moments of Figure \ref{moments}.

While we fit
for $V_c$ and $r_c$ in the cored models, these parameters are strongly 
degenerate.  Our model grid extends to $3.3$ kpc, thus any model with
$r_c > 3.3$ kpc has a uniform density $\rho_c=3V_c^2/4\pi Gr_c^2$ over the
entire range of our model.  Furthermore, we have no velocity information from
stars past 
$R \geq 1.6 \text{ kpc}$ and therefore cannot constrain the kinematics 
in the outer parts of the galaxy.  Thus, for models with 
$r_c \gsim 1.6 \text{ kpc}$,
$\rho_c$ is now the only parameter
that differentiates between models.   As $\rho_c$ is dependent on both
$V_c$ and $r_c$, the latter two parameters are completely degenerate.

Figure \ref{chi2res} illustrates this effect.  Plotted are the $\chi^2$
curves for each model parameter.  Lines of the same color indicate a common 
parameterization of the mass profile (e.g. cored + IMBH).  While the 
$\chi^2$ for both $V_c$ and $r_c$ asymptotes to large values, $\rho_c$
is tightly constrained.  Note that the behavior of $r_c$ for logarithmic
profiles with an IMBH (green line) is a result of incomplete parameter
sampling.  With a more densely-sampled parameter space, the $\chi^2$ curve for
$r_c$ for cored models with an IMBH would likely asymptote to large $r_c$ 
in a similar fashion as models without an IMBH (red curve).

The addition of a central black hole to the mass profile does not
make a noticeable difference to the overall $\chi^2$ for most values of 
\mbh.  We therefore place a $1$-$\sigma$ upper limit on
$M_{\bullet} \leq 3.2 \times 10^4 \, M_{\odot}$.

\begin{figure}[t]
\includegraphics[width=9cm]{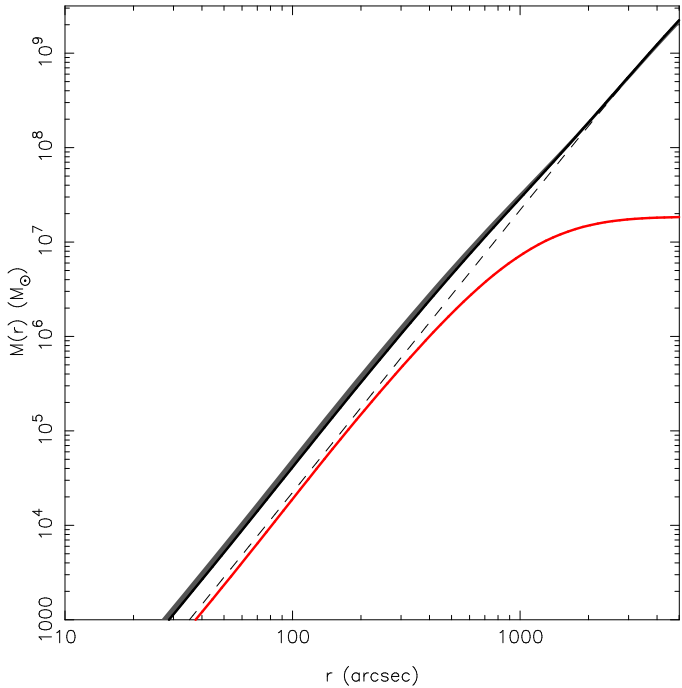}
\caption{Total enclosed mass for our best-fit model (black line with 
surrounding confidence region).  Red line is the enclosed stellar mass.
Dashed line is our best-fit dark matter halo.
\label{massfig}}
\end{figure}

We plot the mass profile for our best-fit model in Figure \ref{massfig}
(solid black line with surrounding $68\%$ confidence region).
This is a cored logarithmic dark matter profile without a central black hole.
The mass profile of our best-fit dark halo is plotted as the dashed line,
and the stellar mass profile is plotted in red.  The contribution of dark 
matter to the total mass increases with radius as the local dynamical 
mass-to-light ratio rises from approximately $\sim 2$ to greater than
100 in the outermost bin of our model.

\subsection{Orbit Structure}

We construct a distribution function for the galaxy from the set of orbital
weights $w_i$ resulting from the $\chi^2$ minimization of our best-fit model.
To explore the orbit structure, we determine the internal (unprojected) 
moments of the distribution function in spherical coordinates.  Streaming 
motions in the $\mathbf{r}$ and $\pmb \theta$ directions are assumed 
to be zero.  In this coordinate system, cross-terms of the velocity dispersion
tensor are zero.

\begin{figure}[t]
\includegraphics[width=9cm]{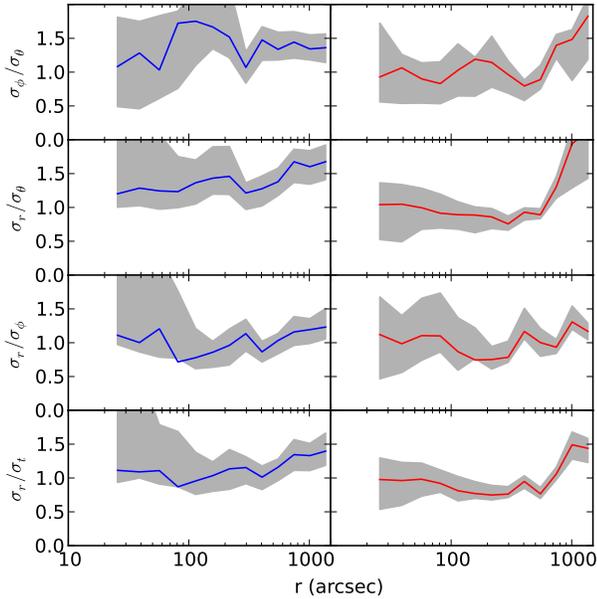}
\caption{Anisotropy in various components of the velocity dispersion tensor.  
Shaded
regions correspond to the 68\% confidence regions, solid lines plot
the best fit model.  Left and right hand panels plot stars near the major and 
minor axes, respectively.
\label{aniso}}
\end{figure}

Figure \ref{aniso} plots the anisotropy in the diagonal components of the
dispersion tensor.  While some panels show an average value near 
unity, there are regions in every panel where the ratio plotted is different
from one.  Additionally, we define the tangential velocity dispersion 
$\sigma_t \equiv \sqrt{\frac{1}{2}(\langle v^2_{\phi} \rangle +
 \sigma^2_{\theta})}$
where $\langle v^2_{\phi}\rangle$ is the second moment 
$\langle v^2_{\phi} \rangle  = \sigma^2_{\phi} + V^2_{\phi}$, and $V^2_{\phi}$ 
is the mean rotation velocity.  With this definition, we plot the ratio
$\sigma_r/\sigma_t$ in the bottom panels of Figure \ref{aniso} 
to investigate whether orbits are radially or tangentially biased.
From these plots it is clear that
the common assumptions of Jeans modeling---constant or zero anisotropy---are
unrealistic.  We find that at most radii in the galaxy, orbits are 
radially biased.  The uncertainty in the anisotropy is largest at small
radii, as evidenced by the size of the 68\% confidence regions in
Figure \ref{aniso}.  This is likely due to the sparsity of kinematics in the
inner part of the galaxy (there are limits to how closely target fibers can
be spaced in multi-fiber spectroscopy).

In a recent
paper, \citet{kaz11} simulated the effects of tidal stirring on a number
of dSph progenitors around a Milky Way sized halo.  They found radial anisotropy
in all of the final remnants, and our models are consistent with these
findings.



\section{Discussion}

\subsection{Cores and Cusps}

Our analysis shows that for the Fornax dwarf an NFW dark matter halo with 
inner slope $\alpha = 1$ is
rejected with high confidence.  We have
kinematics from $30$ pc-$1.6$ kpc, and over this range the models prefer an
$\alpha = 0$ uniform density core with 
$\rho_c = 1.6 \times 10^{-2} \, M_{\odot} \, \mathrm{pc}^{-3}$.  We do not 
attempt to fit for models with an intermediate value of the slope
$0 \le \alpha \le 1$.  Further investigation is necessary before we can
conclude that the best fitting dark matter profile is the logarithmic 
model.  The steep $\alpha=1$ cusp of the NFW profile is, however, robustly
ruled out.  

The models, in general, seem to prefer less mass in the areas over which
we have kinematic constraints.  In NFW models, the concentration $c$ sets the 
normalization (or y-intercept) of the density profile.  Because $c$ cannot be 
lowered below an astrophysically reasonable limit, NFW models enclose more 
mass than cored models.  This difference is reflected in the $\chi^2$ difference
between cored and NFW models, as the kinematics are best fit by models with
less mass.  Figure \ref{moments} hints at this as the best
fit NFW model (dashed line) typically has higher values for $\sigma$ than
either the data or best-fitting cored model (solid line). 

Several groups have approached the core/cusp issue in dSphs by
taking advantage of the fact that some dSphs host multiple
populations of tracer stars that are chemically and dynamically distinct.  
By fitting models to each component, the underlying dark matter profile can
be modeled more accurately.  \citet{amo11b} fit two-component distribution 
function models to Sculptor, while \citet{wal11} apply
a convenient mass estimator (discussed below) to each stellar component
in Sculptor and Fornax.  It is believed that this mass estimator is unaffected
by orbital anisotropy, thus their method 
yields a robust determination of the dynamical mass at two locations in the
galaxy---allowing for the slope of the dark matter profile to be measured.
Each of these studies finds models with a cored dark matter halo 
preferable to the predicted cuspy NFW profile.

It must be noted, however, that we are not observing the pristine 
initial dark matter distribution in this galaxy.  Rather, it has likely been 
modified by 
complex baryonic processes over the lifetime of the galaxy.  These processes
may include: adiabatic compression \citep{blu86}, halo rebounding following 
baryonic mass loss from supernovae \citep{nav96b},
or possibly dynamical friction acting on clumps of baryons
(\citealt{elz01}; but see also \citealt{jar09}).  Although we chose this galaxy
because these effects were likely to be small, they are nevertheless not 
well understood and our result must be taken in that context.

\subsection{Central IMBH}

We are unable to place a significant constraint on the mass of a central
IMBH.  Figure \ref{chi2res} (upper right) shows the marginalized $\chi^2$ curve
against IMBH mass for cored dark matter density profiles.  The curve
asymptotes to low values of IMBH, thus we are only capable of placing an upper
limit on the mass of any potential IMBH.  Furthermore, our best-fit cored model
with and without an IMBH have the same $\chi^2$.  We therefore impose a 
$1$-$\sigma$ upper limit on $M_{\bullet} \leq 3.2 \times 10^4 \, M_{\odot}$.
It is unfortunate that we are not able to place a lower limit on \mbhs
because measurements of black holes in the range \mbh$\lsim 10^4 \, M_{\odot}$
place direct constraints on SMBH formation mechanisms (van Wassenhove et al. 2010 ).  Our models, however,
do robustly rule out a black hole of larger mass.

In massive galaxies it is thought that the radius of influence,
$R_{\mathrm{inf}} \sim G M_{\bullet}/\sigma^2$ must be resolved in order to
detect and precisely measure a black hole \citep{geb03, kor04, fer05,gul09}.
Using our upper limit on $M_{\bullet}$ we can calculate the maximum radius of 
influence of a potential black hole.  Estimating the central velocity 
dispersion at $\sigma \sim 10 \text{ km s}^{-1}$
gives an upper limit for $R_{\mathrm{inf}} \lsim 14$ pc.  Our kinematics start at
$R = 26$ pc, so it is not surprising that the minimum black hole mass we were
able to detect has $R_{\inf}$ close to $ 26$ pc.  
To detect smaller black holes,
we require kinematics of stars closer to the center of the galaxy.

We are able to detect the dynamical influence of a black hole with a similar
mass as \citet{val05} detect in NGC~205, however with kinematics of much lower
resolution.  Our
innermost model bin is centered around $30 \mathrm{~pc}$ whereas they use 
high-resolution kinematics from the \emph{Hubble Space Telescope} 
to resolve spatial scales less than $1 \mathrm{~pc}$.  The advantage we have
is that the central velocity dispersion is much smaller in Fornax, which 
makes $R_{\mathrm{inf}}$ larger for fixed \mbh.  NGC~205 is also more than
five times as distant as Fornax.

\subsection{Mass Estimators}

Several authors have come up with convenient estimators of total mass within
a given radius for local group dSphs.  \citet{stri08} use the mass enclosed
within $300$ pc while \citet{wal09b} and \citet{wol10} find a similar expression
for the mass contained within the projected and un-projected half-light radii,
respectively.
These estimators bear striking resemblance to a result obtained by \citet{cap06}
derived from integral field kinematics of massive elliptical galaxies, and they
all hint at an easy way to determine dynamical masses without expensive 
modeling.  They are believed to be insensitive to velocity anisotropy
based on the derivation in \citet{wol10}, and we compare their estimates
to our models as a check on this.

For the mass contained within $300$ pc we measure
$M_{300}=3.5^{+0.77}_{-0.11} \times 10^6 M_{\odot}$, roughly a factor of three
smaller than \citet{stri08} who measure $M_{300} = 1.14^{+0.09}_{-0.12} \times
10^7 M_{\odot}$ using Jeans models with parameterized 
anisotropy.  

The \citet{cap06}, \citet{wal09b}, and \citet{wol10}
mass estimators are all of the form:

\begin{equation}
M(r_{\mathrm{est}}) = k \langle\sigma^2_{LOS}\rangle R_e
\label{esteq}
\end{equation}

\ni where $r_{\mathrm{est}}$ is the radius at which the estimator is valid.
For \citet{cap06} and \citet{wal09b} $r_{\mathrm{est}} = R_e$
(the projected half-light radius), while for \citet{wol10} 
$r_{\mathrm{est}} = r_e$ (the un-projected half-light radius).
Other than the projected/un-projected difference, each estimator 
differes only by the value of the
constant $k$.  In order to more fairly compare between these estimators and
our models, we use the values for the luminosity-weighted line-of-sight 
velocity dispersion $\langle\sigma^2_{LOS}\rangle=11.3^{+1.0}_{-1.8} \, 
\mathrm{ km~s}^{-1}$
projected half-light radius $R_e=689 \, \mathrm{pc}$, and un-projected 
half-light radius $r_e=900 \, \mathrm{pc}$ that we calculate from the data 
used in our models.

Our best-fitting model has 
$M(R_e)=3.9^{+0.46}_{-0.11} \times 10^7 \, M_{\odot}$
and $M(r_e)=5.8^{+1.0}_{-0.2} \times 10^7 \, M_{\odot}$.
With each group's value for $k$ and our kinematics, 
the mass estimates are: 
$M(R_e) \approx 5.1^{+1.0}_{-1.5} \times 10^7 \, M_{\odot}$ \citep{wal09b}, 
$M(r_e)\approx 8.1^{+1.6}_{-2.4} \times 10^7 \, M_{\odot}$ \citep{wol10}, and
$M(R_e)\approx 1.0^{+0.3}_{-0.2} \times 10^8 \, M_{\odot}$ \citep{cap06}.
Our model is broadly consistent with both the \citet{wal09b} and \citet{wol10} 
estimators.

The evidence that mass estimators are anisotropy-independent
comes largely from comparison to spherical Jeans models (except
\citealt{cap06}). The weakness of
these models is that the anisotropy must be parameterized and is restricted
to be a function of radius only.  Our models are not subject to these 
constraints since the anisotropy is calculated non-parametrically and is
free to vary
with position angle.  We suggest that the best way to prove the accuracy of
mass estimators is to compare with models that can self-consistently 
calculate both mass and anisotropy for realistic potentials.

For bright elliptical galaxies, \citet{cap06} and \citet{tho11} have done
just that.  In these cases, the mass estimates are checked against 
masses derived from axisymmetric Schwarzschild modeling and good agreement
is found.  Ours is the first study to perform a similar test with dSphs, and
there is no reason to assume that success with bright ellipticals guarantees
accuracy in the dSph regime.  The results from our comparison above are
nevertheless reassuring.

\subsection{Tidal Effects}

The principle of orbit superposition, and hence our entire modeling procedure,
relies on the assumption that the galaxy is bound and in a steady state.
The amount of tidal stripping in Fornax due to the effect of its orbit through
the Milky Way's halo is not well-known.  For reasonable values of Fornax
total mass $m$, Milky Way mass $M$, and Galactocentric radius $R_0$, 
the tidal radius of Fornax is $r_t \sim (m/3M)^{1/3} R_0 \sim 13.5 
\text{ kpc}$.  This estimate of $r_t$ is sufficiently larger than our model
grid that we would not expect tidal effects to be important over the radial
range of our models.  If Fornax is on an eccentric orbit about the Milky Way,
however, the above equation for
$r_t$ is not valid and estimation of the tidal radius is not as
straightforward.  Fortunately, studies investigating its transverse motion
suggest the orbit of Fornax is roughly circular \citep{pia07,wal08}.

\begin{acknowledgements}

KG acknowledges support from NSF-0908639.  We thank the Texas Advanced 
Computing Center (TACC) for providing state-of-the-art computing resources.
We are grateful to the Magellan/MMFS Survey collaboration
for making the stellar velocity data publicly available.  Additionally,
we thank Matthew Walker, Mario Mateo, Joe Wolf, and the anonymous referee
for helpful comments on an earlier draft of the paper.

\end{acknowledgements}

\bibliographystyle{apj}

\clearpage

\clearpage

\end{document}